# Hyperspectral Image Reconstruction for Predicting Chick Embryo Mortality Towards Advancing Egg and Hatchery Industry

Md. Toukir Ahmed[1], Md Wadud Ahmed[1], Ocean Monjur[1], Jason Lee Emmert[2], Girish Chowdhary[1,3], Mohammed Kamruzzaman[1]*

[1]Department of Agricultural and Biological Engineering, University of Illinois at Urbana-Champaign, Urbana, IL 61801, USA.

[2]Department of Animal Sciences, University of Illinois at Urbana-Champaign, Urbana, IL 61801, USA.

[3]Department of Computer Science, University of Illinois at Urbana-Champaign, Urbana, IL 61801, USA.

Abstract

As the demand for food surges and the agricultural sector undergoes a transformative shift towards sustainability and efficiency, the need for precise and proactive measures to ensure the health and welfare of livestock becomes paramount. In the context of the broader agricultural landscape outlined, the application of Hyperspectral Imaging (HSI) takes on profound significance. HSI has emerged as a cutting-edge, non-destructive technique for fast and accurate egg quality analysis, including the detection of chick embryo mortality. However, the high cost and operational complexity compared to conventional RGB imaging are significant bottlenecks in the widespread adoption of HSI technology. To overcome these hurdles and unlock the full potential of HSI, a promising solution is hyperspectral image reconstruction from standard RGB images. This study aims to reconstruct hyperspectral images from RGB images for non-destructive early prediction of chick embryo mortality. Firstly, the performance of different image reconstruction algorithms, such as HRNET, MST++, Restormer, and EDSR were compared to reconstruct the hyperspectral images of the eggs in the early incubation period. Later, the reconstructed spectra were used to differentiate live from dead chick-producing eggs using the XGBoost and Random Forest classification methods. Among the reconstruction methods, HRNET showed impressive reconstruction performance with MRAE of 0.0955, RMSE of 0.0159, and PSNR of 36.79 dB. This study motivated that harnessing imaging technology integrated with smart sensors and data analytics has the potential to improve automation, enhance biosecurity, and optimize resource management towards sustainable agriculture 4.0.

**Keywords:** Hyperspectral imaging; Embryo mortality; Agriculture 4.0; Deep learning; Image reconstruction; Classification

1. Introduction

The global surge in per capita egg consumption highlights the urgent need for implementing modern automated egg sorting and grading systems to address the growing demand for eggs across diverse markets. The fourth industrial revolution and global concerns for animal welfare and environmental factors have emphasized the need for non-destructive and non-invasive techniques in the egg and hatchery industry [1,2]. HSI, an emerging fast and accurate non-destructive optical sensing technique, captures the electromagnetic spectrum in hundreds of narrow and contagious bands, offering deep insights into the complexities of egg properties [3,4]. The HSI technique facilitates the visualization of the spatial distribution of quality attributes, providing a powerful tool for real-time monitoring and control in various applications [5–8]. Many studies reported that HSI could have a revolutionary effect on the egg and hatchery industry due to its potential to detect crucial parameters like egg gender, fertility, and embryonic mortality [9,10]. By integrating smart sensing technologies with HSI, hatcheries can be able to comprehensively monitoring of the hatchery environment and make informed decisions such as adjusting incubation conditions, selective breeding practice, and identification of early dead embryos to improve biosecurity (Wang et al., 2023; Yao et al., 2022). The death of developing embryos within eggs, known as embryonic mortality, has substantial consequences on economic viability, productivity, genetic integrity, animal welfare, and environmental sustainability [12]. Embryo mortality may result from genetic defects, environmental stressors, or pathogenic infections [13]. Even though the embryo develops within 24-36 hours (hr) of incubation, chicken eggs typically require 21 days of incubation to hatch [14]. However, non-destructive early detection of chick embryo mortality may allow efficient identification and removal of non-viable embryos, ultimately enhancing hatching rates and improving overall poultry production efficiency.

While HSI offers numerous advantages, its adaption in industrial settings is hindered by high cost, vast data volume, and low data acquisition speeds [15]. The high cost of hyperspectral cameras and associated hardware impedes widespread adaptation of HSI technology, particularly in establishments with limited budgets. Additionally, substantial data volume and complexities necessitate significant investment in computational resources [16]. A critical limitation of HSI in egg research is its reliance on illumination techniques crucial for revealing internal properties, which may disrupt egg biology and pose a risk to embryonic growth [17]. Therefore, unlocking the full potential of HSI beyond its existing limitations, computational approaches such as hyperspectral image reconstruction from conventional RGB images may open new doors in scientific research, industrial applications, and decision-making processes.

Hyperspectral image reconstruction involves extracting information in many contagious bands from standard RGB images with three bands. One of the primary benefits of hyperspectral image reconstruction lies in its accessibility, as it avoids the requirements for expensive specialized HSI equipment. The rising interest in hyperspectral image reconstruction is driven by the user-friendly, widely available nature of RGB imaging, characterized by low hardware costs and minimal data storage requirements [15,18]. However, the reconstruction approach poses significant challenges due to mapping complexities, spectral variability across scenes, spatial-spectral trade-offs, noise, and artifacts in RGB images [19]. Several approaches with advanced architecture, data augmentation, transfer learning, and noise reduction techniques have been reported to overcome the existing challenges in hyperspectral image reconstruction [20]. An appropriate hyperspectral image reconstruction model can significantly help the egg and hatchery industries with various tasks such as grading and sorting, quality monitoring, and early prediction of fertility and gender without disturbing egg biology like conventional HSI systems.

Despite significant advantages, only a few studies have explored hyperspectral image reconstruction for analyzing agricultural and biological samples [21,22]. Initially, researchers reconstructed hyperspectral images from learned manifolds rather than directly from RGB images from an assumption that hyperspectral pixels can be embedded in low-dimensional manifolds [20]. Since the initiation of the New Trends in Images Restoration and Enhancement (NTIRE) challenge focused on spectral reconstruction from RGB images, various convolutional neural networks (CNN), attention, and transformer-based methods have been introduced [23]. These methods hold the potential to effectively reconstruct complex agricultural and biological samples, although their effectiveness has yet to be fully demonstrated. Recently, Yang et al. [22] used the pre-trained MST++ algorithm to develop inversion models for the psychological parameters of rice, with an $R^2$ of 0.40 for predicting the SPAD value. Zhao et al. [21] developed a robust HSCNN-R model for hyperspectral image reconstruction from RGB images to assess the soluble solid content (SSC) of tomatoes, achieving an $R^2$ of 0.51. These studies highlighted the potential of image reconstruction for real-time monitoring of agricultural and biological samples. Building on this foundation, this study marks the first exploration into reconstructing hyperspectral images of eggs to advance egg and hatchery research. More specifically, this study aims to apply advanced reconstruction algorithms, such as HRNET, MST++, Restormer, and EDSR to reconstruct hyperspectral images from RGB images for early prediction of chick embryo mortality.

## 2. Materials and methods

### 2.1 Sample preparation

This study follows the animal experiment protocol (protocol #: 22224) approved by the Office of the Vice-Chancellor for Research IACUC online protocols, the University of Illinois Urbana-Champaign (UIUC). A total of 102 white-shell chicken eggs were collected from the UIUC poultry farm. All the mother birds were White Leghorn breeds at approximately 29 weeks of age. The eggs were incubated in a single-stage commercial incubator (Natureform Inc., Jacksonville, FL, USA) at 99.5º F, 85% relative humidity, with 45º automatic hourly rotation. At the end of the complete incubation period, egg samples were broken out to check the embryo mortality.

### 2.2 Image acquisition

A line-scan benchtop HSI system (Resonon Inc., Bozeman, MT, USA) was used for hyperspectral image acquisition (Fig. 1). The system had a 900-pixel line-scanner camera (Pika L, Resonon, Inc.) in the range of 400-1000 nm with a lens of 23 mm focal length, a wide-spectrum illumination system with tungsten-halogen lamp, a linear movable translation stage, and a computer. Turning off the room lights, all scans were performed using Spectronon Pro software (Resonon Inc., Bozeman, MT, USA) from the computer connected to the camera system. The individual sample was placed vertically in a black holder with the air cell up, and the light was mounted just beneath the sample. Before imaging, the light source was turned on for 30 min to ensure stable illumination throughout the experiment. To maintain uniform image quality, the distance from the camera lens to the sample surface (20 cm) and acquisition parameters in the Spectronon Pro software, including "stage" (scanning speed: 0.07 cm/s, homing speed: 1.5625 cm/s, jogging speed: 0.07 cm/s) and "imager" (frame rate: 9.9 fps, integration time: 100 ms, gain: 0), remains constant. Before image acquisition, each egg case (30 eggs) was removed from the incubator and kept in an electric blanket (90-95º F) until the image acquisition of the batch. Following the same protocol, hyperspectral images were acquired from 2 to 4 days of incubation. Therefore, a total of 306 images (102×3 days) were obtained for consecutive analysis. It should be noted that when light is incident on an egg sample, it scatters in multiple directions instead of passing straight through, distorting its path and intensity. Such light scattering causes some light to be absorbed or reflected by the egg, complicating the separation of transmitted light. Thus, this study used digital number values [24] (also known as pixel intensity) instead of transmittance or absorbance spectra for subsequent analysis.

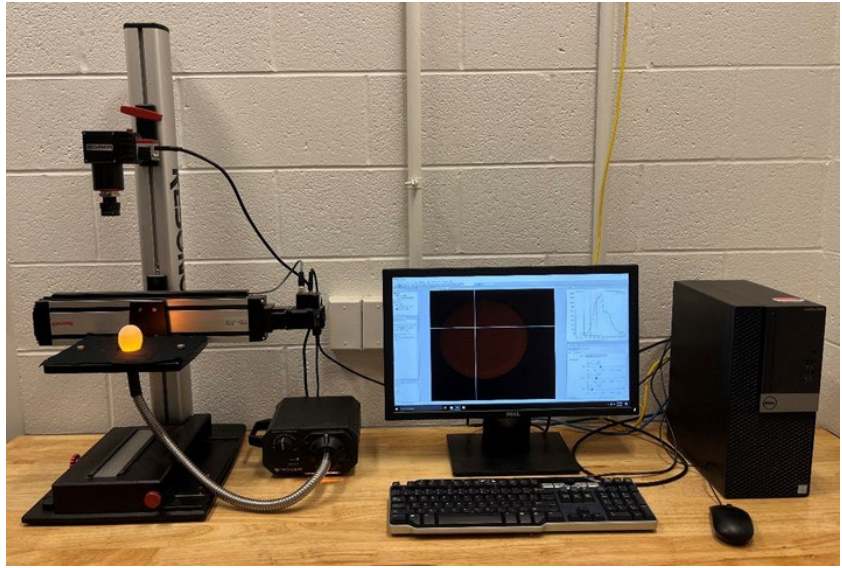

Fig. 1. The line-scan hyperspectral imaging system for non-destructive egg analysis.

2.3 Image segmentation and extraction of spectral data

Image segmentation is an essential initial step in extracting spectral information from hyperspectral images [7]. It involves identifying regions of interest (ROI) within the tested sample by creating a mask that distinguishes them from the background. In this study, the mask was generated at 700 nm (the band with the greatest contrast difference between the sample and background) with a threshold of 0.02. This mask subsequently served as the ROI for extracting spectral information from the hyperspectral images. A single mean spectrum was derived for each hyperspectral image by calculating the mean of all pixel spectra within each ROI to derive the spectra from the ROIs.

2.4 Data preparation for reconstruction methods

The reconstruction algorithms require RGB images as inputs and use hypercubes for label data. However, the Resonon HSI system does not produce RGB images along with the hyperspectral images. To address this issue, pseudo-RGB images were created using specific wavelengths - 600 nm (red), 549 nm (green), and 450 nm (blue). These images were then saved in the JPEG (Joint Photographic Experts Group) format, as illustrated in Fig. 2. These pseudo-RGB images then served as the inputs for the reconstruction algorithms.

The ground-truth hypercubes were constructed by selecting random non-collinear wavelengths from the visible to near-infrared (VIS-NIR) spectrum, covering the range of 400 to 1000 nm. This selection process resulted in each hypercube with dimensions of 800×900×$\lambda$, where $\lambda$ represents the number of selected wavelengths. These hypercubes were then used as the label data to train the hyperspectral image reconstruction algorithms.

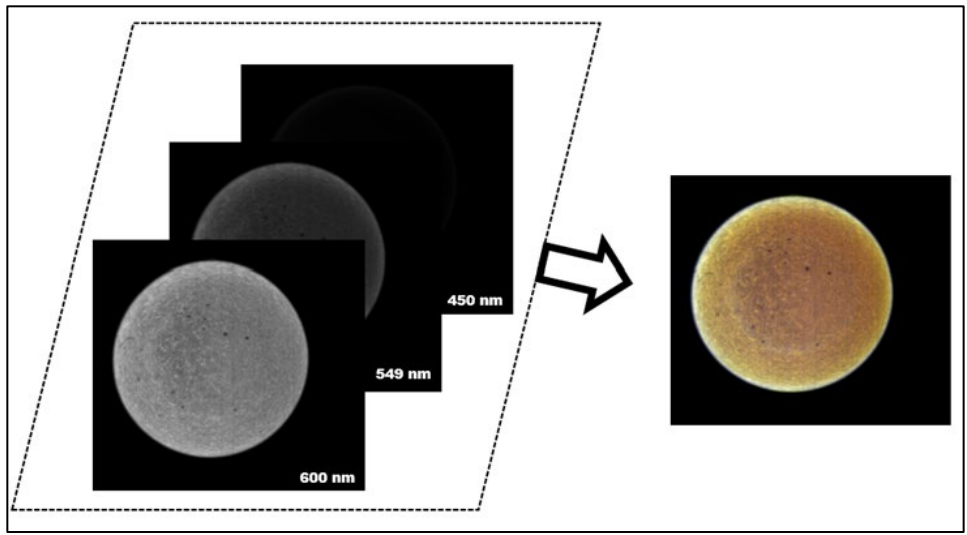

Fig. 2. Generation of pseudo-RGB images combining 600, 549, and 450 nm bands.

2.5  Reconstruction algorithms

Four models with different architectures were selected to provide a fair performance comparison for the reconstruction task (Fig. 3a-d). EDSR is CNN-based, while Restormer and MST++ primarily use attention in their respective architectures. HRNet is predominantly a CNN-based architecture with some self-attention elements in the structure. Convolutions and attention-based networks focus on different attributes of an image. Attention networks act as low-pass filters compared to convolutions, which are high-pass [25]. By benchmarking complementary architectures, this study aims to identify possible architectural directions to take for hyperspectral image reconstruction in eggs.

HRNet breaks down input images in hierarchies by downsampling them by factors of two [26]. HRNet incorporates PixelUnshuffle to downsample the input images without adding parameters. These hierarchies are processed along with the upsampled features of the immediate lower hierarchy. The upsampling is done using PixelShuffle. This process cascades up to the topmost layer for the final reconstruction output. The upsampling is learned instead of using fixed methods. HRNet tries to provide a way to prevent the loss of information while downsampling.

MST++ uses a Single-Stage spectral-wise Transformer (SST), which is composed of multiple mask-guided spectral-wise multi-head self-attention modules [27]. The masks attend to more important features of the hyperspectral representation. Allowing MST++ to model the long-range intra-spectral dependencies of hyperspectral images accurately.

Restormer, primarily a transformer-based image restoration model, is also used in hyperspectral image reconstruction [28]. Restormer uses self-attention across the feature dimensions rather than the spatial dimensions, making it ideal for learning dependencies across different channels. The transformer blocks in the restormer architecture use a multi-Dconv head transposed attention. This attention layer performs query-key feature interaction across channels. Its proposed gated feed-forward network allows only informative features to propagate across the network.

Based on the SRResNet [29], EDSR [30] stacks multiple residual blocks for hyperspectral reconstruction. The authors found that by removing the batch normalization layers in the residual blocks, the performance does not degrade substantially, but the memory usage does. This allows EDSR to stack more residual blocks with the same memory usage.

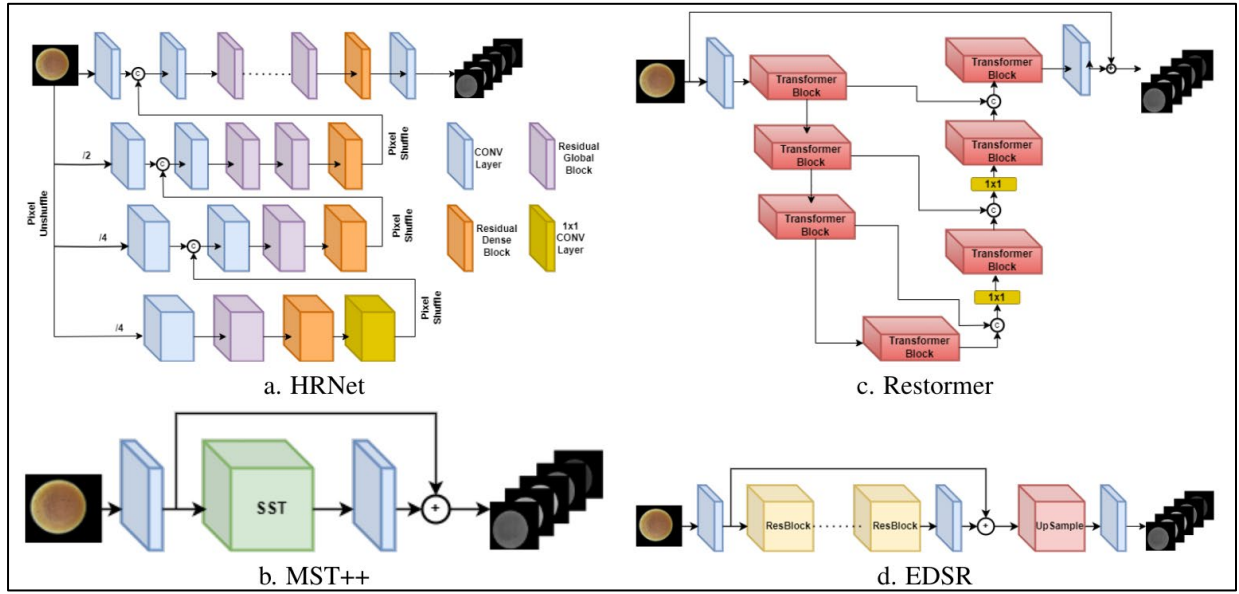

Fig. 3. Architectures of HSI reconstruction models. All the reconstruction models take RGB as input and provide HSI (λ bands) as output.

## 2.6 Reconstruction performance evaluation metrics

### 2.6.1 Mean Relative Absolute Error (MRAE)

MRAE calculates the relative absolute error between the predicted and ground truth (GT) labels. By measuring the average relative difference, MRAE is scale-invariant compared to other metrics, which take only the absolute difference.

$$MRAE = \frac{1}{N}\sum_{i=1}^{N} \frac{|Pred^i - GT^i|}{GT^i} \qquad (1)$$

### 2.6.2 Root mean squared error (RMSE)

RMSE takes the squared error of each prediction label and its associated GT and takes the root of it. This makes RMSE more prone to outliers, allowing its use in cases where outliers are weighted heavily.

$$RMSE = \frac{1}{N}\sum_{i=1}^{N}\sqrt{(Pred^i - GT^i)^2} \qquad (2)$$

### 2.6.3 Peak signal-to-noise ratio (PSNR)

PSNR measures how well the reconstructed image preserves the original information. It represents the ratio between the maximum possible power of the signal compared to the noise.

$$PSNR = 10 log_{10}\left(\frac{R^2}{MSE}\right) \qquad (3)$$

## 2.7 Spectral classification

Given the small sample size of the spectral dataset, XGBoost[31] and Random Forest [32] were chosen for classification. XGBoost is a boosting algorithm that sequentially adds weak learners and makes an overall ensemble where each weak learner heavily focuses on errors made by the previous weak learners. XGBoost is effective for datasets where overfitting is a concern. On the other hand, Random Forest is an ensemble of decision trees. Each decision tree in a Random Forest has random sets of features and trains on a random subset of the data. For the final classification result, the class with the most votes from the decision trees is the output. The classification methods used in this study were evaluated using the accuracy, precision, recall, and F1 score (Eq. 4-7).

$$Accuracy = \frac{TP+TN}{TP+FP+TN+FN} \qquad (4)$$

$$Precision = \frac{TP}{TP+FP} \tag{5}$$

$$Recall = \frac{TP}{TP+FN} \tag{6}$$

$$F1\ Score = 2 \times \frac{Precision \times Recall}{Precision+Recall} \tag{7}$$

Where *TP* is true positive, *FP* is false positive, *TN* is true negative, and *FN* is false negative.

### 2.8 Computational environment

The computations in this study were carried out using Google Colaboratory Pro (Colab Pro), a cloud computing platform. The Colab Pro virtual machine was powered by a two-core Intel(R) Xeon(R) CPU at 2.30GHz, 25 GB of RAM, and an extended lifetime, as well as an NVIDIA Tesla P100 GPU with 16 GB VRAM. The analyses were carried out using Python 3.9. The reconstruction models were trained using open-source Python packages, including Scikit-learn, PyTorch, and OpenCV.

## 3. Results and Discussion

### 3.1 Training the reconstruction models

The first step in training the reconstruction models involved processing input RGB and label hyperspectral data. To prepare the hyperspectral data used as labels, 10 non-collinear bands (520, 583, 619, 655, 700, 739, 780, 837, 870, and 903 nm) were chosen within the VIS-NIR spectrum (400-1000 nm). These bands were selected based on spectral analysis where spectral differences were prominent for live and dead embryos. The chosen bands are mainly related to blood formation, yolk dissolution, water, protein, and lipid change during embryonic development [33–35]. Changes in these components reflect the dynamics of interactions between the chick embryo and the egg composition [36,37]. The feature reduction process was applied to eliminate redundant multi-collinear information from the original hyperspectral images, ensuring that the model's predictive accuracy on biological products remained unaffected [38]. The processed hyperspectral images, with the selected bands, were used as ground-truth labels, whereas the corresponding RGB images served as inputs for the models. The dataset consisted of 306 images, which were randomly divided into training (80%), validation (10%), and testing (10%) sets. The training of the models employed the Adam optimizer, with adjustments to the decay rates $\beta_1=0.9$ ($\beta_1=0.5$ for HRNet) and $\beta_2=0.99$. Training parameters were set to a batch size of 1, a patch size of 128, and a stride of 8. The initial learning rate was established at $1e^{-4}$, featuring a systematic exponential decrease with each epoch to optimize the training process.

After setting all the hyperparameters, each model was trained for 200 epochs, each involving 1000 iterations. The MRAE loss function was chosen for its ability to evaluate wavebands under various lighting conditions uniformly and its high resistance to outlier data, as noted by Shi et al. [39]. Furthermore, using the MRAE loss function led to models achieving quicker convergence within fewer epochs, according to Zhao et al. [21]. During training, the loss stabilized after several epochs, indicating no further significant improvements. Notably, the best-performing models were selected for further evaluation based on their performance before the loss plateaued; this occurred at different epochs for each model: HRNET at the 189th epoch, MST++ at the 175th, Restormer at the 180th, and EDSR at the 195th. These models were then assessed using the validation and test sets. The outcomes of this evaluation are detailed in Table 1.

Table 1: Evaluation metrics of the hyperspectral image reconstruction models on validation and test set. Bold indicates the best performance in respective metrics.

| Methods | MRAE (val) | RMSE (val) | MRAE (test) | RMSE (test) | PSNR (dB) |
|---|---|---|---|---|---|
| HRNet | **0.0936** | **0.0151** | **0.0955** | **0.0159** | **36.79** |
| MST++ | 0.2615 | 0.0378 | 0.2615 | 0.0377 | 13.51 |
| Restormer | 0.1111 | 0.0188 | 0.1111 | 0.0188 | 35.32 |
| EDSR | 0.1136 | 0.0178 | 0.1136 | 0.0178 | 35.70 |

HRNet stood out among the reconstruction techniques, achieving the best results with the lowest MRAE at 0.0936 and RMSE at 0.0151, along with the highest PSNR at 36.79 dB. However, it was also the least efficient, requiring 40.5 hr and consuming 9.49 gigaflops (G) of computational resources. On the other hand, MST++ was the most efficient in terms of computational resources (0.15 G) and time (12.5 hr), but it lagged behind in PSNR performance, recording only 13.51 dB for the reconstructed images. The Restormer technique, while being relatively efficient computationally (5.39 G), took longer to complete (37 hr) but achieved a commendable PSNR of 35.32 dB. Lastly, the EDSR method outperformed both MST++ and Restormer in PSNR, reaching 35.70 dB, despite having higher computational demand at 9.17 G.

### 3.2 Visual quality assessment of the reconstructed images

After the successful execution, a comparative analysis was conducted to evaluate the visual quality of the reconstructed hyperspectral images against the original ground-truth images. Fig. 4 presents both sets of images at selected spectral wavelengths for comparison. Visual inspection of Fig. 4 shows that the reconstructed images closely resemble the ground-truth images. In particular, images produced by the HRNet and Restormer demonstrated a remarkable visual similarity to the ground truth. However, minor discrepancies were observed in the pixel intensities of images reconstructed by Restormer, especially in the 739 and 780 nm wavelengths.

On the other hand, EDSR showed a contrast deviation from the ground truth for the 700 nm bands as marked in Fig. 4, though it showed good similarity for the rest of the bands. Conversely, images reconstructed by the MST++ method exhibited noticeable differences in pixel intensity compared to the ground-truth images, struggling with accurately reconstructing pixel intensities, especially in the 583-700 nm range, as shown in Fig. 4.

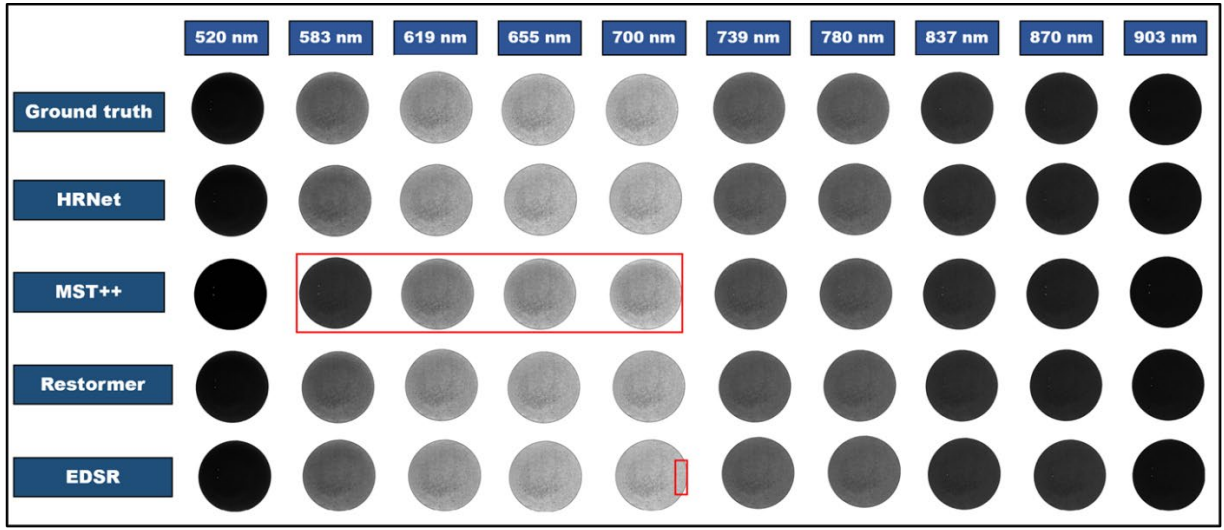

Fig. 4. Ground-truth and reconstructed images. The red area represents the inconsistencies in the reconstructed pixels. MST++ showed the most inconsistencies for reconstructing 583, 619, 655, and 700 nm bands.

3.3 Reconstructed spectral quality assessment

After the image reconstruction process, spectra were extracted from the reconstructed images, and the quality of these spectral reconstructions was evaluated. Fig. 5a-d displays the comparison between the average actual (ground truth) spectra and the reconstructed spectra. HRNet and Restormer achieve the closest matches compared to the actual spectra, whereas EDSR and MST++ exhibit some discrepancies. HRNet, in particular, demonstrates reconstructions that are almost perfectly symmetrical to the actual spectra across all bands, a result anticipated due to HRNet's superior reconstruction capabilities (as depicted in Fig. 5a). Restormer (Fig. 5c) exhibits slight discrepancies in the wavelength ranges of 650 to 700 nm and 740 to 780 nm. EDSR's reconstructions deviate primarily in the 600 to 780 nm range, although they align closely with the actual spectra in other bands (Fig. 5d). MST++ faces challenges in accurately reconstructing most wavebands, indicating its lower performance in reconstruction tasks compared to the other methods (as shown in Fig. 5b). The analysis of spectral quality reveals a clear trade-off between computational complexity and performance. HRNet and Restormer, which offer better performance, are also more computationally intensive than the other methods. Furthermore, the architecture of these methods, especially HRNet's combination of convolutional and attention layers, likely allowed for a more accurate understanding of pixel patterns compared to the other methods.

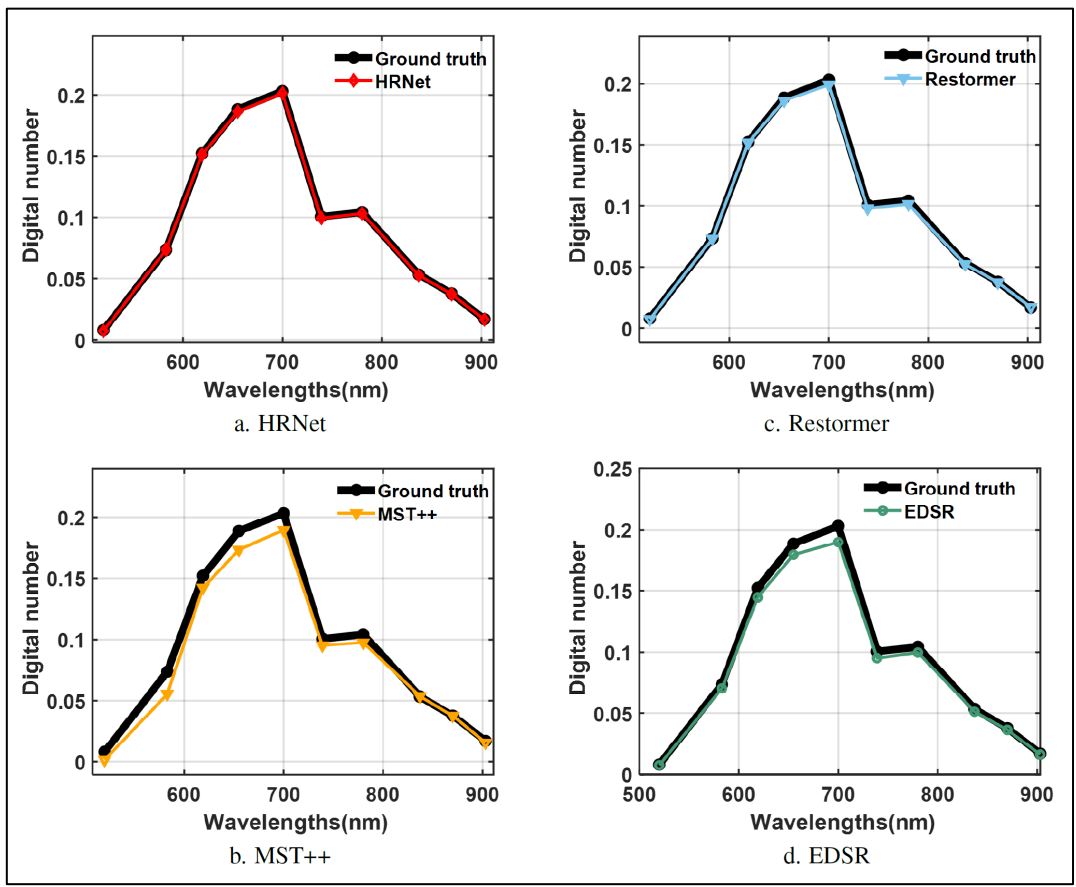

a. HRNet    c. Restormer
b. MST++    d. EDSR

Fig. 5. Comparison of average ground truth and reconstructed spectra of egg samples. Spectra extracted from HRNet and Restormer showed the most similarity with their ground truth counterparts, while MST++ struggled to reconstruct most of the bands.

## 3.4 Classification based on the reconstructed spectra

With the same reference value (i.e., embryo mortality), only single-day samples (# 102) were utilized to evaluate classification performance. Up to day 2, all samples exhibit nearly identical spectral properties, while significant structural and compositional changes manifest from day 3 onwards [40]. Thus, spectra extracted from day 3 images were selected for applying classification models. It is noteworthy that among the 102 samples, only 9 showed deceased embryos, with embryonic death attributed to uncontrolled factors such as genetic defects and environmental stressors. For the classification algorithms, to account for this highly imbalanced class distribution, SMOTE was applied to oversample the minority class. Using SMOTE, synthetically generated data for dead embryo classes were added to the classification dataset. Stratified 5-fold cross-validation was used to test the classification dataset. The overall classification results in Table 2 validate the previously acquired spectral reconstruction results.

HRNet had the lowest deviation from the ground truth for spectral reconstruction. Similarly, Table 2 shows that HRNet is the best-performing reconstruction model, with accuracy, precision, recall, and F1 scores of 81.15, 87.26, 75.20, and 79.19% using Random Forest. The classification performance obtained from the reconstructed spectra of Restormer gives the second-best results among the reconstructed models, while reconstructed spectra from EDSR and MST++ give the worst classification performance.

Table 2: Comparative classification results for predicting live and dead embryos. The blue-colored results indicate the best possible performance obtained from the ground truth, while the bold-colored results show the best possible performance of the reconstruction models.

| Spectra | Classification | Accuracy | Precision | Recall | F1 Score |
|---|---|---|---|---|---|

|  | methods | (%) | (%) | (%) | (%) |
|---|---|---|---|---|---|
| Ground truth | XGBoost | 83.31 | 87.28 | 79.64 | 83.01 |
|  | Random Forest | 84.93 | 89.33 | 79.59 | 83.96 |
| HRNet | XGBoost | 80.62 | 83.60 | 76.31 | 79.66 |
|  | Random Forest | **81.15** | **87.26** | 75.20 | 79.19 |
| Restormer | XGBoost | 80.64 | 82.85 | **78.47** | **80.20** |
|  | Random Forest | 80.62 | 84.81 | 75.20 | 79.42 |
| EDSR | XGBoost | 76.34 | 78.38 | 72.98 | 75.23 |
|  | Random Forest | 77.42 | 78.95 | 75.26 | 76.26 |
| MST++ | XGBoost | 73.12 | 73.62 | 72.10 | 72.82 |
|  | Random Forest | 74.69 | 78.66 | 68.77 | 73.09 |

## 4. Conclusions

Hyperspectral image reconstruction has the prospect of becoming a popular domain for transforming agri-vision research by developing cost-effective and smart solutions. This approach can unlock the full potential of hyperspectral imaging by using consumer-level RGB images. Realizing this potentiality, for the first time, this study demonstrated the suitability of hyperspectral image reconstruction for chick embryo mortality prediction during the early incubation period. This study used HRNET, MST++, Restormer, and EDSR algorithms to reconstruct hyperspectral images from standard RGB images. With MRAE, RMSE, and PSNR of 0.0955, 0.0159, and 36.79 dB, respectively, the HRNet model outperformed all other reconstruction methods used in the study, showing remarkable similarities with the ground truth images and spectra. Consequently, XGBoost and Random Forest techniques were applied to differentiate samples of live and dead embryos using spectra of reconstructed images. Aligned with the reconstruction performance observed, the Random Forest classification on the reconstructed spectra from the HRNet model gave the best results. These encouraging findings inspire continued research into cost-effective solutions for hatchery operations, advancing toward a more efficient and sustainable industry that embraces the principles of Industry 4.0.